\documentclass[11pt]{article}

\usepackage{acl}

\usepackage{times}
\usepackage{latexsym}
\usepackage{amssymb}
\usepackage{amsmath}
\usepackage{cleveref}
\newtheorem{problem}{Problem}
\usepackage{booktabs, tabularx, makecell}
\usepackage{multirow}
\usepackage{adjustbox}

\usepackage[T1]{fontenc}

\usepackage[utf8]{inputenc}

\usepackage{microtype}

\usepackage{inconsolata}
\usepackage[most]{tcolorbox}
\newtheorem{hypothesis}{Hypothesis}
\usepackage{graphicx}
\usepackage{booktabs}
\usepackage{algorithm}
\usepackage{algpseudocode}
\usepackage{tikz}
\usetikzlibrary{positioning,arrows.meta,fit,backgrounds,shapes.geometric}

\newcommand{\ours}{EquivAug}

\title{Diverse Reasoning Paths Matter: Symbolic Equivalence Augmentation for Enhancing RL Exploration
}

\author{
 \textbf{Xiaoou Liu\textsuperscript{1}},
 \textbf{Tiejin Chen\textsuperscript{1}},
 \textbf{Hanyang Chen\textsuperscript{1}},
 \textbf{Yuan Ling\textsuperscript{2}},
 \textbf{Hua Wei\textsuperscript{1}},
\\
 \textsuperscript{2}Independent Researcher, 
 \texttt{ericalingyuan@gmail.com} \\
 \textsuperscript{1}Arizona State University,
 \texttt{\{xiaoouli, tchen169, hchen478, hua.wei\}@asu.edu} \\
}

\begin{document}
\maketitle
\begin{abstract}

Reinforcement learning with verifiable rewards (RLVR) has emerged as a powerful paradigm for improving the reasoning capabilities of LLMs. However, the role of supervised fine-tuning (SFT) data in enabling effective downstream RL remains underexplored. In this paper, we investigate a key hypothesis: \emph{the diversity of reasoning steps in chain-of-thought (CoT) during SFT is a critical enabler of RLVR performance}. To test this hypothesis, we introduce \textbf{\ours}, a data augmentation framework that uses symbolic equivalence transformations to generate diverse yet provably correct reasoning paths. By converting CoT solutions into formal symbolic representations, applying equivalence transformations, and reconstructing natural-language explanations, \ours{} produces training data that differs in reasoning steps rather than surface-level lexical variation. Experiments on GSM8K and FOLIO across model scales demonstrate that \ours{} consistently achieves the highest reasoning path diversity after SFT and the strongest accuracy after GRPO\footnote{Our code can be found at \url{https://github.com/Xiao0o0o/EquivAug}.}. Our results establish that CoT diversity during SFT is positively correlated with downstream RLVR performance.

\end{abstract}

\section{Introduction}

Most prior work on supervised fine-tuning (SFT) for reasoning focuses primarily on improving accuracy on the SFT task itself~\citep{yuan2023scaling}. However, achieving high SFT accuracy does not necessarily lead to a better model, especially after the success of reinforcement learning with verifiable rewards (RLVR)~\cite{guo2025deepseek,singh2025openai}. Recent studies show that high SFT accuracy does not reliably lead to high RL performance, but enabling exploration is more important~\cite{kang2025quagmires}. However, existing work has controlled for variables at the problem-level factors, such as task diversity~\cite{kang2025quagmires}, which affect RL performance, a fundamentally different dimension remains underexplored: the \emph{reasoning path level}.  This raises a natural question: \emph{Does exposure to diverse chain-of-thought during SFT enable more effective exploration and stronger performance in downstream RLVR?}

 In this paper, we propose a novel perspective: \textbf{the diversity of reasoning steps in chain-of-thought (CoT) at the SFT stage is a critical enabler of RLVR}. A model fine-tuned on a single solution per problem may achieve strong supervised performance, yet it learns a rigid policy that RL struggles to effectively diversify. We hypothesize that exposing the model to \emph{multiple structurally distinct yet correct} reasoning paths during SFT broadens the initial policy's support, thereby enabling more effective exploration during the subsequent RL phase. Despite its intuitive appeal, to the best of our knowledge, no prior work has systematically investigated how the diversity of SFT reasoning data influences downstream RLVR performance.

To systematically analyze the relationship between reasoning diversity during the SFT stage and performance after RLVR, we must first address a closely related challenge: \emph{how to obtain diverse reasoning paths}. Existing data augmentation methods, such as paraphrasing-based augmentation ~\citep{yu2024metamath,li2024mugglemath,li2024neurosymbolic} and Rejection Sampling Fine-Tuning (RFT)~\cite{yuan2023scaling}, typically rely on LLMs to generate alternative phrasing for a single solution. However, these approaches primarily introduce surface-level lexical variation rather than new reasoning strategies. Therefore, current data augmentation methods cannot provide the diversity of reasoning steps in CoT and thus are not suitable to verify our hypothesis.


To address the limitation of current data augmentation methods and study the relationship between diversity and RL, we introduce \textbf{\ours}, a novel data augmentation framework that generates diverse and verified reasoning paths. Our approach is grounded in a key observation: mathematical and logical reasoning inherently admits rich equivalence structures, in which the same conclusion can be reached through multiple formally equivalent derivations. For example, in mathematics, one can apply the distributive property to alter the computational graph (e.g., transforming $a(b+c)$ to $ab+ac$). By employing symbolic equivalent transformations, \ours{} ensures that the augmented chain-of-thought (CoT) trajectories are structurally distinct at intermediate steps. 

We conduct experiments on the GSM8K \citep{cobbe2021training} and FOLIO \citep{han2024folio} datasets, evaluating our method across model scales ranging from 1.5B to 8B parameters. Using Group Relative Policy Optimization (GRPO)~\citep{shao2024deepseekmath} as our RL algorithm, we compare different data augmentation strategies followed by an identical post-training process. 
Our results validate our core hypothesis: introducing diverse CoT data during the SFT stage significantly enhances post-training RLVR performance, while surface-level diversity cannot. This highlights the vital importance of ensuring the \emph{reasoning step diversity} during the SFT stage. Overall, our main contributions can be summarized as follows:

\begin{itemize}

    \item We systematically investigate the relationship between CoT diversity at the SFT stage and downstream RLVR performance. We demonstrate that diverse and correct SFT data significantly improve RL outcomes. To the best of our knowledge, this is the first work to empirically establish this connection.
    \item We propose \ours, a novel framework that generates diverse and verifiable reasoning paths via equivalent transformations. 
    \item We empirically show that \ours, combined with GRPO, consistently outperforms all baseline methods across multiple model scales (1.5B – 8B) and benchmark. Furthermore, we provide detailed ablation studies on transformation complexity and data scaling.
\end{itemize}

\section{Related Work}
\label{sec:related}

\subsection{The Interplay between SFT and Reinforcement Learning}

Enhancing exploration has been considered the key to RL training~\cite{song2025outcome}. Considering the SFT-then-RL pipeline~\cite{guo2025deepseek,shao2024deepseekmath}, SFT provides a cold-start reasoning policy, which decides the quality of exploration~\cite{chu2025sft,chen2025step}. Consequently, a growing body of work has analyzed how SFT decides the exploration and the following RL performance. For instance, \citet{yuedoes} demonstrates that SFT on data from stronger LLMs can broaden the model's exploration ability. \citet{matsutani2026rl} shows that SFT can expand the number of correct trajectories and therefore increase the exploration ability. \citet{kang2025quagmires} demonstrates that high accuracy after SFT does not lead to high accuracy after RL, showing that task diversity or difficulty are more important in deciding performance after RLVR.  However, while these studies highlight the importance of SFT data, they primarily analyze factors at the \emph{problem level} such as diversity of questions, ignoring the diversity of reasoning paths. In contrast, we investigate the \textit{reasoning path level}: for a fixed set of problems, how the diversity of chain-of-thought solutions shapes the model's initial exploration space and thereby influences downstream RL optimization to fill the gap.

\subsection{Data Augmentation for Reasoning}

Data augmentation has become a central strategy for improving the reasoning capabilities of LLMs during SFT. A lot of work focuses on increasing the \emph{complexity and diversity of questions}: WizardMath~\citep{luo2023wizardmath} applies evolutionary instructions to increase problem difficulty, MetaMath and MuggleMath~\citep{yu2024metamath,li2024mugglemath} bootstraps mathematical questions by rewriting them from multiple perspectives. 
For augmenting the \emph{solution side}, existing approaches fall into two broad categories. \emph{Paraphrase-based} methods employ a stronger model to rewrite ground-truth solutions~\cite{yu2024metamath,li2024neurosymbolic}. These methods increase surface-level diversity but preserve the underlying solution strategy. \emph{Sampling-based} methods, exemplified by Rejection Sampling Fine-Tuning (RFT)~\citep{yuan2023scaling}, use the model itself to generate multiple candidate solutions and retain those whose final answer is correct. However, final-answer correctness does not guarantee the validity of intermediate reasoning steps, because LLMs can produce plausible but logically flawed chains of thought.
A promising direction to address these limitations is \emph{neuro-symbolic} data augmentation, which leverages symbolic representations to ensure the correctness of generated data. Recent work has explored symbolic mutation for generating verifiable math problems~\citep{li2024neurosymbolic} and symbolic verification for logical reasoning data~\citep{li2024gsm}. Our work builds on this line of research: \ours{} combines equivalence-preserving symbolic transformations with solver-based verification to produce reasoning paths that are both structurally diverse and provably correct at every intermediate step.

\section{Motivation and Problem Definition}

In GRPO~\cite{shao2024deepseekmath}, the policy is updated based on non-zero advantages, which are computed by normalizing rewards within a group of $G$ sampled responses to the same prompt:
\begin{equation}
    \hat{A}_i = \frac{r_i - \text{mean}(\{r_j\}_{j=1}^{G})}{\text{std}(\{r_j\}_{j=1}^{G}) + \epsilon},
\end{equation}
where $r_i = R(x, y_i)$ is the binary reward for response $y_i$ and $\epsilon$ is a small constant for numerical stability. These advantages provide meaningful learning signals only when the group contains responses with different rewards. Since the rewards in RLVR are mostly rule-based accuracy rewards~\cite{xie2025logic}, if all $G$ responses are uniformly correct or incorrect, the advantages will vanish, leaving the policy with no gradient for that prompt.

Therefore, for an effective RL process, the policy $\pi_{\theta_{\text{SFT}}}$ must be capable of producing sufficiently diverse responses so that sampled groups contain a mixture of correct and incorrect solutions~\cite{zhang2025edge}. This requires exploration ability for $\pi_{\theta_{\text{SFT}}}$. A policy trained on narrow SFT data tends to generate homogeneous responses, increasing the risk of uninformative reward groups. Moreover, on challenging problems, a narrow policy may fail to produce any correct response~\cite{wu2026learn}. A broader initial policy, equipped with multiple reasoning strategies, is more likely to succeed on such problems through at least one strategy, reducing the occurrence of all-incorrect groups and ensuring that informative advantage signals persist throughout training. The initial policy could learn from SFT data to obtain multiple reasoning strategies, which motivates our following hypothesis:

\begin{hypothesis}
\label{hyp:diversity}
Training the model on diverse reasoning paths during SFT broadens the initial policy's coverage of the solution space, leading to more effective exploration and stronger downstream RLVR performance.
\end{hypothesis}

To test Hypothesis~\ref{hyp:diversity}, we need SFT datasets with diversity on CoT. However, existing augmentation methods such as paraphrase-based rewriting~\citep{yu2024metamath,you2024mumath} primarily introduce surface-level lexical variation that changes the phrases or sentences without altering the underlying reasoning structure. Training on such data does not broaden the reasoning strategies available to the policy, making it unsuitable for testing our Hypothesis~\ref{hyp:diversity}. This leads to the central technical problem of our work: how can we generate multiple reasoning paths that are step-level diverse while preserving correctness? Formally, we define this as the following augmentation problem:

\begin{problem}[Equivalent Solution Augmentation]\label{prob:equiv-aug}

Given a dataset $\mathcal{D} = \{(q_i, s_i, a_i)\}_{i=1}^{N}$, where $q_i$ is a question, $s_i$ is a natural language reasoning path, and $a_i$ is the final answer, the goal is to produce an augmented dataset that $i$-th sample has $K_i$ reasoning paths:
\[
\mathcal{D}' = \left\{(q_i,\; s'_{i,k},\; a_i) \mid k = 1, \ldots, K_i \right\},
\]
where each generated reasoning path $s'_{i,k}$ satisfy two key constraints:
 (1)~\emph{correctness}: $s'_{i,k}$ stick to the same answer $a_i$ and 
 (2)~\emph{diversity}: two variants $\sigma'_{i,k}$ and $\sigma'_{i,k'}$ have different reasoning paths.
\end{problem}

\section{Methodology}
\label{sec:method}

In this section, we introduce \ours, a framework for symbolic data augmentation that uses equivalence transformations to generate diverse yet provably correct reasoning paths. We first describe the overview and then detail each component.

\begin{figure*}
    \centering
    \includegraphics[width=\linewidth]{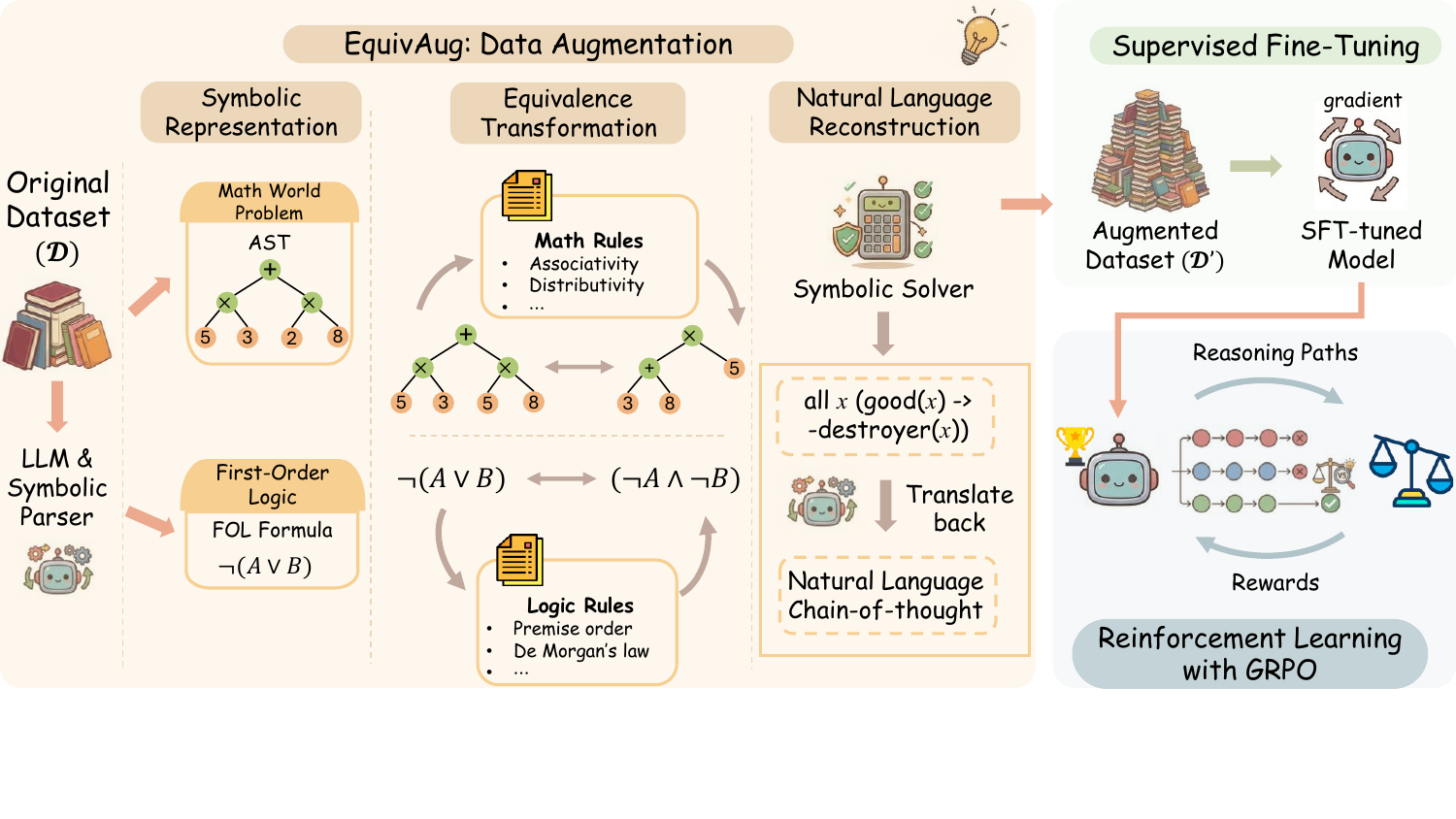}
    \vspace{-20mm}
    \caption{ Overview of our pipeline. \textbf{Left:} The \ours{} data augmentation framework converts chain-of-thought solutions into symbolic representation, applies equivalence transformation rules to generate structurally diverse yet semantically equivalent variants verified by a symbolic solver, and reconstructs them back into natural language. \textbf{Right:} The augmented data is used for supervised fine-tuning, followed by reinforcement learning with GRPO.}
    \label{fig:overview}
    \vspace{-5mm}
\end{figure*}

\subsection{Overview}

As illustrated in the left portion of Figure~\ref{fig:overview}, \ours{} is a domain-agnostic augmentation framework comprising three modules: (1) \textbf{Symbolic Representation}, which converts natural language CoT into formal symbolic forms. Specifically, math word problems are translated into algebraic expressions, while logical reasoning problems are encoded as solver-compatible first-order logic formulas; (2) \textbf{Equivalence Transformation}, which applies domain-specific equivalence rule libraries to systematically derive multiple alternative yet semantically equivalent symbolic solutions from each original representation; and (3) \textbf{Natural Language Reconstruction}, which converts each derived symbolic solution back into natural language, producing the final augmented training dataset.

Crucially, every augmented CoT solution is verified by a symbolic solver prior to back-translation, ensuring the correctness of not only the final answer but also all intermediate reasoning steps. This symbolic design distinguishes \ours{} from heuristic augmentation methods that risk introducing subtle errors. The entire pipeline is modular: the same three-stage workflow applies uniformly across different domains such as mathematics and logic, differing only in the domain-specific equivalence rule sets plugged into Stage~(2). Appendix~\ref{sec:algorithm} summarizes the complete procedure.


\subsection{Symbolic Representation}

The first module maps each chain-of-thought solution into a formal symbolic representation that supports both symbolic verification and equivalence-preserving transformation. While the specific representation differs across domains, the underlying principle remains the same: the original reasoning process is converted into a structured symbolic object that can be parsed, manipulated, and verified by external solvers. We detail the domain-specific instantiations below.

\noindent\textbf{Math Word Problems.} We decompose the solution process into a sequence of step-level algebraic expressions and organize them into a unified abstract syntax tree (AST)~\cite{alfred2007compilers}. In this tree, leaf nodes correspond to numbers and variables, while internal nodes denote arithmetic or algebraic operators such as $+$, $-$, $\times$, and $\div$. To ensure correctness, we execute the complete symbolic expression using a symbolic solver and compare the resulting value against the gold answer; only solutions that pass this verification are retained. This tree-structured formulation enables the equivalence-preserving transformations listed in Table~\ref{tab:math-rules} to be applied to arbitrary subtrees.

\noindent\textbf{First-Order Logic.} For first-order logic (FOL) reasoning problems, the premises and conclusions in the FOLIO dataset~\cite{han2024folio} are already annotated as FOL formulas. To obtain a solver-executable symbolic representation, we convert these formulas into the input format required by automated theorem provers via lightweight rule-based parsing.

We employ the automated theorem prover Prover9~\cite{mccune2005release} to verify the entailment labels and extract structured proof traces. Prover9 adopts a refutation-based proving strategy: it negates the target conjecture, adds it to the premise set, and attempts to derive a contradiction. Depending on the gold label (\textsc{True}, \textsc{False}, or \textsc{Uncertain}), we configure the conjecture accordingly and invoke either Prover9 or the finite model builder Mace4 to verify the label. Details about Prover9 can be found in Appendix~\ref{app:prover9}.

A key property of Prover9 that our framework exploits is that its proof search is highly sensitive to both the syntactic form of the input formulas and the search strategy configuration. Even semantically equivalent reformulations of the premises, such as reordering clauses, applying De Morgan's laws, or distributing quantifiers, can lead the prover down substantially different search paths, yielding structurally distinct yet equally valid proof traces. This sensitivity provides a natural mechanism for generating diverse reasoning paths: by systematically rewriting the input formulas using semantics-preserving logical equivalences, we can obtain multiple alternative derivations from a single problem instance without compromising logical validity.

\subsection{Equivalence Transformation}
\label{sec:equivalence-transformation}

The second module generates diverse symbolic variants by systematically applying equivalence-preserving transformations to each formal representation. These rewrites alter the structure of the symbolic reasoning process while ensuring that every transformed variant remains consistent with the original answer or logical label. We maintain domain-specific transformation libraries, summarized for mathematics and first-order logic in Table~\ref{tab:math-rules} and Table~\ref{tab:folio-rules}, respectively.

\noindent\textbf{Math Word Problems.} We apply local rewrites to the symbolic expression tree using algebraic equivalence rules. These rewrites operate on subtrees of the expression AST and include standard properties such as commutativity, associativity, distributivity, and fraction decomposition, listed in Table~\ref{tab:math-rules}. Because each rewrite is applied to a well-formed subexpression, the transformed expression remains mathematically equivalent to the original one. 
These transformations can change the decomposition order of arithmetic operations, the grouping structure of intermediate expressions, and the representation of fractions or chained operations, while leaving the final numerical value unchanged. 
By sampling different valid rewrites over the symbolic tree, we obtain multiple correct symbolic solution variants from a single original CoT.

\noindent\textbf{First-Order Logic.}
For FOL reasoning, we transform the symbolic premises using logical equivalences, including double negation, contraposition, De Morgan's laws, implication expansion, and related rewrites~\cite{enderton2001mathematical}. Since premise order does not affect logical entailment, we also apply premise-level permutations such as shuffling and reordering. The full set of logical transformations is listed in Table~\ref{tab:folio-rules}.
Although these rewrites preserve the logical content of the original premises, they often cause the theorem prover to explore different inference paths and produce distinct proof traces. As a result, multiple transformed premise sets may correspond to the same gold label while exhibiting substantially different derivational structures.

Beyond formula-level transformations, we also increase reasoning diversity by varying the search configuration of the theorem prover. Specifically, we use multiple Prover9 strategies that differ in resolution settings and clause-selection heuristics, including hyperresolution, unit-resulting resolution, and different values of \texttt{pick\_given\_ratio}. These strategy variations do not alter the semantics of the input formulas; instead, they change how the prover searches for a refutation.


\begin{table}[t]
\centering
\scriptsize
\setlength{\tabcolsep}{3pt}
\begin{adjustbox}{width=\linewidth}
\begin{tabular}{@{}llp{3.5cm}@{}}
\toprule
Category & Rule & Equivalence \\
\midrule
\multirow{6}{*}{Arithmetic}
& Commutativity &
\makecell[l]{
$a+b \leftrightarrow b+a$ \\
$a \times b \leftrightarrow b \times a$
} \\
& Associativity &
\makecell[l]{
$(a+b)+c \leftrightarrow a+(b+c)$ \\
$(a \times b)\times c \leftrightarrow a \times (b \times c)$ \\
$a \times b \div c \leftrightarrow a \div c \times b$
} \\
& Distributivity &
\makecell[l]{
$a \times (b+c) \leftrightarrow a \times b + a \times c$ \\
$(a+b)\times c \leftrightarrow a \times c + b \times c$ \\
} \\
& Distributivity over subtraction &
\makecell[l]{
$a \times (b-c) \leftrightarrow a \times b - a \times c$
} \\
\midrule
\multirow{3}{*}{Fractions}
& Decomposition and composition &
\makecell[l]{
$(a+b)/c \leftrightarrow a/c + b/c$ \\
$(a-b)/c \leftrightarrow a/c - b/c$ \\
$a/b + c/b \leftrightarrow (a+c)/b$ \\
$a/b - c/b \leftrightarrow (a-c)/b$
} \\
& Multiplication/division restructuring &
\makecell[l]{
$a \div b \div c \leftrightarrow a \div (b \times c)$ \\
$(a/b)/c \leftrightarrow a/(b \times c)$ \\
$a \times (b/c) \leftrightarrow (a \times b)/c$ \\
$(a/b)\times c \leftrightarrow (a \times c)/b$ \\
$a/(b/c) \leftrightarrow (a \times c)/b$
} \\
& Reordering &
\makecell[l]{
$a \times b \div c \leftrightarrow a \div c \times b$
} \\
\midrule
Subtraction & Chain conversion &
\makecell[l]{
$a-b-c \leftrightarrow a-(b+c)$
} \\
\bottomrule
\end{tabular}
\end{adjustbox}
\caption{\small Equivalence rules for \textbf{math world problems}.}
\label{tab:math-rules}
\end{table}

\begin{table}[t]
\centering
\scriptsize
\setlength{\tabcolsep}{3pt}
\begin{adjustbox}{width=\linewidth}
\begin{tabular}{@{}llp{4.0cm}@{}}
\toprule
Category & Rule & Equivalence / Transformation \\
\midrule
\multirow{5}{*}{Premise order}
& Original &
\makecell[l]{Keep the original premise order}
\\
& Shuffle &
\makecell[l]{Permute premises arbitrarily}
\\
& Reverse &
\makecell[l]{Reverse the premise order}
\\
& Rotate &
\makecell[l]{Cyclically rotate the premise order}
\\
& Pairwise swap &
\makecell[l]{Swap adjacent premise pairs}
\\
\midrule
\multirow{6}{*}{Logical equivalence}
& Double negation &
\makecell[l]{
$A \leftrightarrow \neg\neg A$
}
\\
& Contraposition &
\makecell[l]{
$A \rightarrow B \leftrightarrow \neg B \rightarrow \neg A$
}
\\
& De Morgan's law &
\makecell[l]{
$\neg(A \land B) \leftrightarrow \neg A \lor \neg B$ \\
$\neg(A \lor B) \leftrightarrow \neg A \land \neg B$
}
\\
& Implication expansion &
\makecell[l]{
$A \rightarrow B \leftrightarrow \neg A \lor B$
}
\\
& Biconditional expansion &
\makecell[l]{
$A \leftrightarrow B \leftrightarrow (A \rightarrow B) \land (B \rightarrow A)$
}
\\
& Alternative biconditional &
\makecell[l]{
$A \leftrightarrow B \leftrightarrow
(A \land B) \lor (\neg A \land \neg B)$
}
\\
\midrule
\multirow{3}{*}{Connective restructuring}
& Commutativity of $\land,\lor$ &
\makecell[l]{
$A \land B \leftrightarrow B \land A$ \\
$A \lor B \leftrightarrow B \lor A$
}
\\
& Distributivity &
\makecell[l]{
$A \land (B \lor C) \leftrightarrow (A \land B) \lor (A \land C)$ \\
$A \lor (B \land C) \leftrightarrow (A \lor B) \land (A \lor C)$
}
\\
& Antecedent exportation &
\makecell[l]{
$(A \land B) \rightarrow C \leftrightarrow A \rightarrow (B \rightarrow C)$
}
\\
\midrule
\multirow{2}{*}{Simplification}
& Idempotent law &
\makecell[l]{
$A \land A \leftrightarrow A$ \\
$A \lor A \leftrightarrow A$
}
\\
& Absorption &
\makecell[l]{
$A \lor (A \land B) \leftrightarrow A$ \\
$A \land (A \lor B) \leftrightarrow A$
}
\\
\midrule
\multirow{5}{*}{Composite transformation}
& Shuffle + contraposition &
\makecell[l]{
Shuffle premises, then apply contraposition
}
\\
& Shuffle + implication expansion &
\makecell[l]{
Shuffle premises, then expand implications
}
\\
& Shuffle + De Morgan &
\makecell[l]{
Shuffle premises, then apply De Morgan's law
}
\\
& Contraposition + expansion &
\makecell[l]{
Apply contraposition, then implication expansion
}
\\
& De Morgan + double negation &
\makecell[l]{
Apply De Morgan's law with selective double negation
}
\\
\bottomrule

\end{tabular}

\end{adjustbox}
\caption{\small Equivalence transformations for \textbf{first-order logic}.}
\label{tab:folio-rules}
\end{table}

\subsection{Natural Language Reconstruction}

The final module converts each transformed symbolic reasoning path back into natural language CoT solutions. We prompt a large language model to translate the symbolic derivation into fluent natural language explanations. The prompt is carefully designed to enforce strict alignment between the symbolic steps and the generated text. The full prompt template is provided in Appendix~\ref{app:prompt}.


\section{Experiments}
\label{sec:experiments}

We conduct comprehensive experiments to validate the effectiveness of \ours{} and to investigate how diverse yet provably correct reasoning paths during SFT influence both the fine-tuned model itself and its subsequent RL training. Specifically, our experiments are organized around the following research questions:

\noindent\textbf{RQ1:}  Can \ours{} generate diverse CoT paths, and how do other reasoning augmentation methods perform on generating diverse CoT?

\noindent\textbf{RQ2:} Does exposure to diverse chain-of-thought paths during SFT facilitate more effective exploration in post-training reinforcement learning?

\noindent\textbf{RQ3:} How do the complexity and quantity of equivalence transformations affect downstream performance? Can \ours{} generalize to out-of-distribution tasks?

\subsection{Experiment Settings}

\textbf{Datasets.} We evaluate on two benchmarks: (1) GSM8K~\cite{cobbe2021training} is a grade-school math word problems, consisting of 1,319 test examples. (2) FOLIO~\cite{han2024folio} is a first-order logic reasoning benchmark where each example comprises a set of natural language premises paired with FOL annotations and a conclusion to be classified as \textsc{True}, \textsc{False}, or \textsc{Uncertain}.

\paragraph{Configurations.} We experiment with three backbone large language models of varying scale: Qwen2.5-1.5B, Qwen2.5-3B~\citep{yang2025qwen3}, and Llama-3.1-8B~\citep{grattafiori2024llama}. For supervised fine-tuning, we apply full-parameter training for Qwen2.5-1.5B and Qwen2.5-3B and use LoRA~\cite{hu2022lora} for Llama-3.1-8B. We deliberately select models of different sizes from distinct model families to examine whether our findings generalize across model scales and architectures. Full SFT runs use a cosine learning rate schedule with a peak learning rate of 2e-5, while LoRA uses a learning rate of 2e-4. For reinforcement learning, we adopt GRPO~\citep{shao2024deepseekmath} as the post-training algorithm. The reward signal is based on exact-match correctness of the final answer. We use a sampling temperature of 0.7, generate 8 rollouts per prompt with a learning rate of 1e-6. All experiments are conducted on 2 Nvidia A100 GPUs. To have a fair comparison, we use the same training samples for all methods. Further implementation details are provided in Appendix~\ref{app:config}.

\paragraph{Compared Methods.} We compare several reasoning augmented strategies while keeping all other post-training configurations fixed: 

\noindent(1) \textbf{Ground Truth}: SFT on ground-truth solutions. We directly use the annotated solutions provided by the benchmark. For FOLIO, since the benchmark didn't provide the reasoning traces, we directly use the gold label as the ground-truth reasoning. 

\noindent(2) \textbf{RFT}~\citep{yuan2023scaling}: In the Rejection Sampling Fine-tuning method, we use a model that has been fine-tuned on the ground-truth data to sample multiple candidate solutions, retaining only those whose final answers are correct as the augmented training dataset.

\noindent(3) \textbf{Paraphrasing}: We use a stronger language model (e.g., GPT-4o-mini) to paraphrase the original annotated solutions in the benchmark, to make sure the augmented solutions that remain correct while increasing diversity in expression. In the case of FOLIO, the benchmark only provides final labels rather than full reasoning paths. We therefore first use the model itself to generate a chain-of-thought that arrives at the correct answer, and then paraphrase this generated solution with the stronger model. Prompt templates are deferred to Appendix~\ref{app:prompt}.

\noindent(4) \textbf{\ours}: Our method employing the equivalence transformation libraries and symbolic verification pipeline described in Section~\ref{sec:method}.

\textbf{Metrics.} To evaluate the effectiveness of \ours{} and validate our hypothesis, we report accuracy at two stages: \emph{ACC\textsubscript{SFT}}, the final-answer accuracy of the model after SFT, and \emph{ACC\textsubscript{GRPO}}, the accuracy after subsequent reinforcement learning with GRPO. We use mean accuracy for 5 runs. To quantify the diversity of reasoning paths generated by the fine-tuned model, we adopt \textbf{Reasoning Path Diversity} (RPD)~\citep{ju2025reasoning}, a step-level metric that captures semantic variation across intermediate reasoning steps rather than surface-level lexical differences. For each test instance, we sample $5$ solutions from the SFT model and compute RPD over the set of generated reasoning paths. Further details on the RPD computation are provided in Appendix~\ref{sec:app_rpd}.

\subsection{Diversity of Generated Paths (RQ1)}

We first evaluate whether supervised fine-tuning on \ours{} data elicits greater diversity in the model's generated chain-of-thought reasoning compared to baseline methods. This question is central to our framework, since the purpose of symbolic transformation is not only to preserve correctness, but also to expose the model to a broader set of valid reasoning trajectories during training, thereby unlocking greater exploration potential for subsequent reinforcement learning. As shown in Table~\ref{tab:main_table}, models trained with \ours{} achieve higher RPD than the corresponding baselines in nearly all evaluation settings. It can show that symbolic equivalence-based augmentation effectively broadens the space of reasoning patterns learned during supervised fine-tuning.


Notably, the comparison with other baselines is particularly informative: while paraphrasing by GPT-4o introduces lexical variation, it largely preserves the original reasoning structure, resulting in lower RPD. Similarly, RFT improves diversity to some extent by sampling multiple solutions from the model itself, but the generated paths tend to cluster around the patterns already learned during initial SFT, limiting the effective diversity gain. An apparent exception is RFT on Qwen2.5-1.5B for FOLIO, which achieves an RPD of 0.4998. Closer inspection shows that this high score mainly results from unstable output styles, alternating between short direct answers and lengthy premise restatements, rather than genuine reasoning diversity. In contrast, \ours{} produces paths that differ at the level of reasoning steps, such as restructuring intermediate computations or deriving conclusions through alternative inference chains, yielding substantially higher diversity. This suggests that systematic symbolic transformations, rather than mere surface-level rephrasing, are the key driver of reasoning diversity.



\begin{table*}[t!]
\centering
\begin{adjustbox}{max width=\textwidth}
\begin{tabular}{c|ccc|ccc|ccc}
\hline
LLM
& \multicolumn{3}{c|}{\textbf{Qwen2.5-1.5B}}
& \multicolumn{3}{c|}{\textbf{Qwen2.5-3B}}
& \multicolumn{3}{c}{\textbf{Llama3.1-8B}} \\
\cline{2-10}
Metrics
& RPD & ACC\textsubscript{SFT} & ACC\textsubscript{GRPO}
& RPD & ACC\textsubscript{SFT} & ACC\textsubscript{GRPO}
& RPD & ACC\textsubscript{SFT} & ACC\textsubscript{GRPO} \\
\hline

\multicolumn{10}{c}{Dataset: GSM8K} \\
\midrule
Ground Truth & \underline{0.1132}  & 55.88 & \underline{76.87} & 0.1560  & 63.59 & \underline{83.00} & 0.0952  & 62.08 &	72.47 \\
RFT          & 0.0815  & 55.66	& 74.37 & 0.1087  & 67.61 & 80.66 & 0.0619 & 66.35 &	71.87\\
Paraphrasing & 0.0318 & \textbf{62.29}	& 74.52 & \underline{0.1925} & \textbf{75.31} & 82.90 & \underline{0.1051}  & \textbf{72.22} &	\underline{75.20} \\
EquivAug     & \textbf{0.1200} & \underline{61.71}	& \textbf{76.95} & \textbf{0.2087} & \underline{73.36} & \textbf{83.09} & \textbf{0.1102} & \underline{67.81} &	\textbf{76.35} \\
\midrule

\multicolumn{10}{c}{Dataset: FOLIO} \\
\midrule
Ground Truth & 0.0001  & 44.83	& 64.08 & 0.0001 & \textbf{69.56} &	\underline{75.61} & 0.0001  & 64.24 &	73.39 \\
RFT          & \textbf{0.4998}  & 49.75	& \underline{64.97} & \underline{0.1338}  & 60.30 & 68.97 & 0.0878 & \underline{69.56} &	74.28 \\
Paraphrasing & 0.1280 & \textbf{61.58}	& 62.56 & 0.1255  & 65.02 &	72.51 & \underline{0.1213}  & 65.71 &	\underline{75.17} \\
EquivAug     & \underline{0.1720}  & \underline{56.06} & \textbf{65.41} & \textbf{0.1745} & \underline{67.49} & \textbf{78.27} & \textbf{0.1718} & \textbf{70.94} & \textbf{75.61} \\
\hline
\end{tabular}
\end{adjustbox}
\caption{\small Results on GSM8K and FOLIO across different backbone models. We report RPD, ACC\textsubscript{SFT}, and ACC\textsubscript{GRPO} for each data augmentation method. \ours{} produces more diverse reasoning paths after SFT and achieves the best performance after GRPO across all settings.}
\label{tab:main_table}
\vspace{-3mm}
\end{table*}

\subsection{Post-training RL with Diverse SFT (RQ2)}

We then analyze how different SFT training datasets affect downstream RL performance. As shown in Table~\ref{tab:main_table}, models fine-tuned with \ours{} consistently achieve the highest ACC\textsubscript{GRPO} across all model scales and datasets, confirming that equivalence-based augmentation provides the strongest foundation for post-training RL. More importantly, we highlight two key observations:

\noindent\textbf{Reasoning diversity correlates with RL performance.} We observe a clear positive relationship between RPD after SFT and accuracy after GRPO: methods that produce higher reasoning diversity during SFT tend to yield stronger RL performance. For instance, on Qwen2.5-1.5B, \ours{} achieves both the highest RPD, followed by Ground Truth, which ranks second in both diversity and post-RL accuracy
To quantify this relationship, we compute the Spearman rank correlation~\citep{zar2005spearman} between RPD and ACC\textsubscript{GRPO} rank across all methods for each dataset, averaged over model scales. \textbf{The resulting correlations are $\rho = 0.87$ on GSM8K and $\rho = 0.67$ on FOLIO, with an overall average of $\rho = 0.77$}. This strong positive correlation substantiates our central Hypothesis~\ref{hyp:diversity}: training on diverse reasoning paths during SFT broadens the initial policy's coverage of the solution space, facilitating more effective exploration and stronger downstream RL performance.

\noindent\textbf{SFT accuracy does not predict RL accuracy.} Interestingly, \ours{} is not consistently the best method at the SFT stage. Its ACC\textsubscript{SFT} often falls below the strongest baseline. However, after GRPO, \ours{} surpasses all baselines across settings. We attribute this to the nature of the cross-entropy training objective used in SFT: fine-tuning on more diverse reasoning paths increases the entropy of the model's output distribution, making it appear less confident and thus less accurate under greedy decoding at the SFT stage. Yet this higher entropy is precisely what benefits RL, as it preserves a broader policy distribution from which the RL algorithm can explore and optimize. We provide empirical evidence for this by visualizing the entropy trajectories of different models during GRPO training in Appendix~\ref{sec:appendix}.
This finding reveals that the primary advantage of \ours{} lies not in higher immediate supervised accuracy, but in better compatibility with RL-based optimization. Even when this advantage is not fully reflected in SFT accuracy, it becomes evident once the model is further optimized with GRPO.


\subsection{Ablation and OOD Generalization (RQ3)}
To further analyze the relationship between diversity and RL performance, we conduct several additional experiments on \ours{}. More results on entropy and accuracy can be found in Appendix~\ref{sec:more_experiment}.

\noindent\textbf{Amount of augmented data.}
We investigate how the number of augmented reasoning paths affects downstream RL performance. Specifically, we vary the maximum number of augmented solutions per problem ($K \in \{3, 5, 10, 20\}$) and train Qwen2.5-1.5B on GSM8K. As shown in Table~\ref{tab:abl_num_solutions}, both RPD and downstream accuracy peak at $K=5$. Further increasing the number of solutions reduces RPD without improving performance, indicating that the benefit of augmentation depends on the diversity of the generated reasoning paths rather than the sheer amount of augmented data. Based on this result, we use a maximum of five solutions per problem in all GSM8K experiments.


\begin{table}[t]
\centering
\begin{adjustbox}{width=0.9\linewidth}
\begin{tabular}{c|c|cc}
\toprule
\makecell[c]{Max solutions \\ per problem}
& RPD
& ACC\textsubscript{SFT}
& ACC\textsubscript{GRPO} \\
\midrule
3  & 0.1213          & 0.6118          & 74.22          \\
5  & \textbf{0.1234} & \textbf{0.6171} & \textbf{75.06} \\
10 & 0.1187          & 0.5991          & 74.38          \\
20 & 0.1161          & 0.6068          & 73.09          \\
\bottomrule
\end{tabular}
\end{adjustbox}
\caption{Effect of the maximum number of solutions per problem on reasoning diversity and performance (GSM8K on Qwen2.5-1.5B).}
\label{tab:abl_num_solutions}
\end{table}

\begin{figure}
    \centering
    \includegraphics[width=0.7\linewidth]{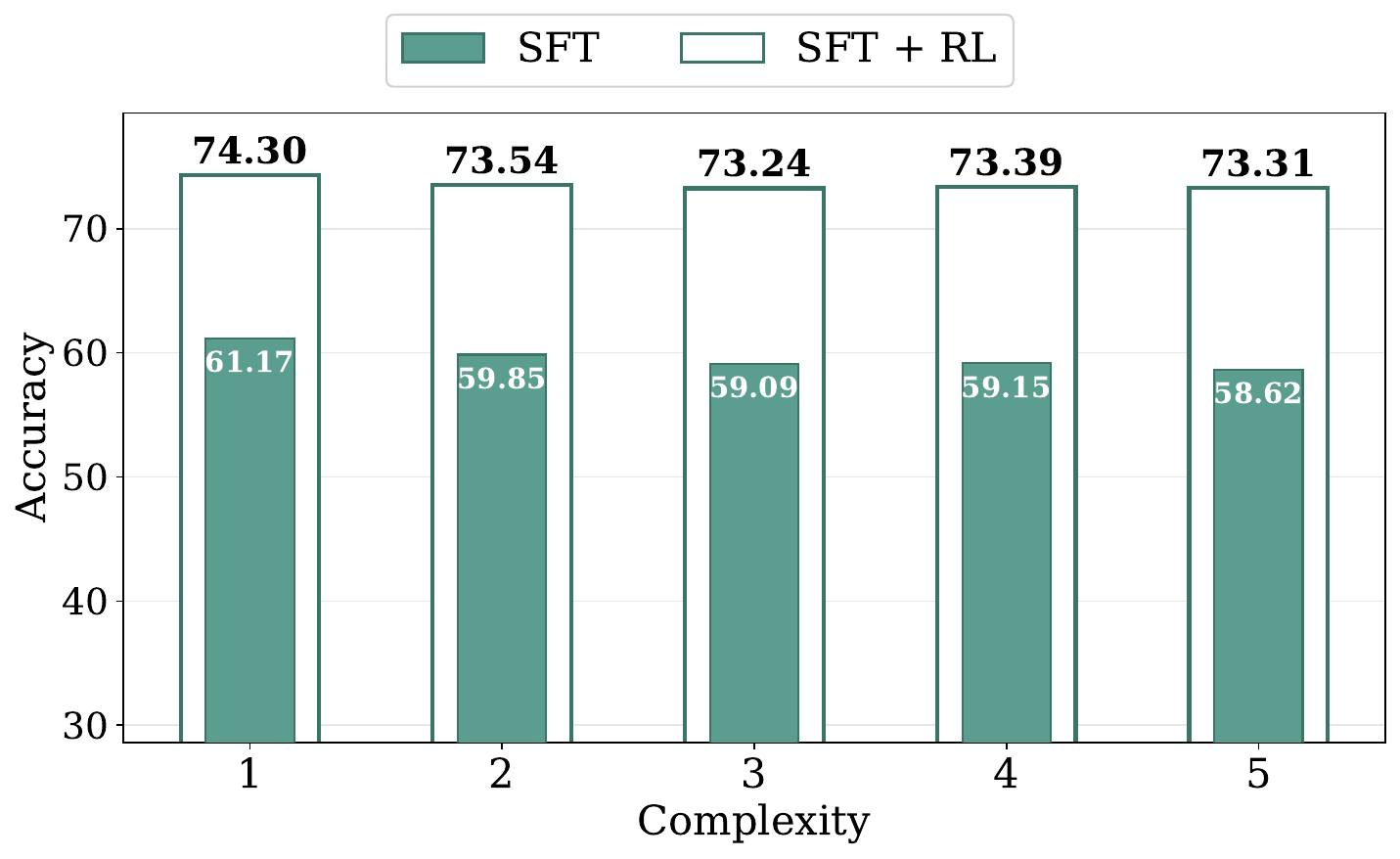}
    \caption{Effect of transformation complexity on performance (GSM8K on Qwen2.5-1.5B).}
    \vspace{-3mm}
    \label{fig:complexity}
\end{figure}

\noindent\textbf{Complexity of transformations.} Next, we examine whether applying more transformation rules to each solution leads to better performance. We conduct an experiment by controlling the transformation complexity, defined as the number of times transformation rules are sequentially applied to an already modified reasoning path on GSM8K and Qwen2.5-1.5B. The relationship between transformation complexity and downstream performance is illustrated in Figure~\ref{fig:complexity}. The results show that a complexity level of 1 yields the best performance. Applying too many sequential transformations degrades accuracy, likely because heavily manipulated reasoning paths become unnatural or structurally strange, which confuses the policy.

\begin{figure}
    \centering
    \includegraphics[width=\linewidth]{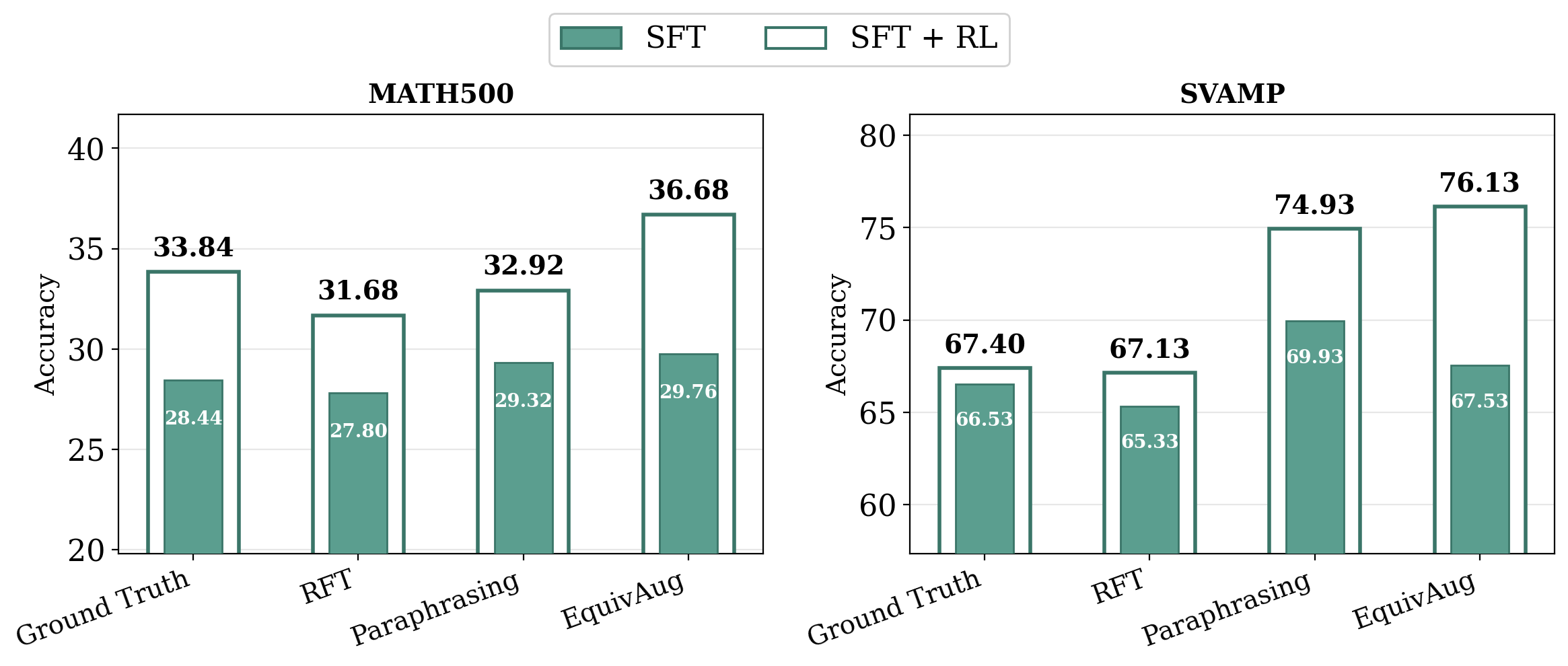}
    \caption{Out-of-distribution performance of different data augmentation methods on MATH500 and SVAMP.}
    \vspace{-4mm}
    \label{fig:ood}
\end{figure}

\noindent\textbf{OOD Generalization.}  Finally, We evaluate whether the diversity introduced by \ours{} transfers to out-of-distribution settings. Using Qwen2.5-1.5B trained on GSM8K, we evaluate on SVAMP~\citep{patel2021nlp}, a distribution-shifted arithmetic reasoning benchmark, and MATH500~\citep{lightman2023lets}, which contains more challenging mathematical problems. As shown in Figure~\ref{fig:ood}, \ours{} achieves the best post-RL accuracy on both datasets, reaching 76.13 on SVAMP and 36.68 on MATH500. These results indicate that the diverse reasoning strategies acquired during SFT generalize beyond the training distribution and remain beneficial across reasoning tasks of different difficulty levels.

\section{Conclusion}

In this work, we investigated the relationship between chain-of-thought diversity during supervised fine-tuning and downstream reinforcement learning with verifiable reward performance. We proposed \ours, a symbolic data augmentation framework that generates structurally diverse yet provably correct reasoning paths through equivalence-preserving transformations and solver-based verification. Experiments on GSM8K and FOLIO across multiple model scales consistently demonstrate that \ours{} produces more diverse reasoning paths after SFT and achieves stronger performance after GRPO. Our findings establish that reasoning-step-level diversity, rather than surface-level lexical variation, is the key factor in preparing an effective policy foundation for RLVR. 
We hope this work encourages further exploration into how the structural properties of SFT data can be designed to better support post-training reinforcement learning.

\section*{Limitations}

This work has two main limitations. First, the equivalence rule libraries used in \ours{} are manually designed based on well-known mathematical and logical identities. While these rules are sufficient to produce meaningful diversity in the current evaluation domains, automatically discovering, learning, or expanding domain-specific equivalence rules is a promising direction that we leave for future work. Second, our experiments are conducted on models ranging from 1.5B to 8B parameters. Although the benefits of reasoning diversity are consistent across these scales, it remains an open question whether these gains persist, diminish, or amplify as model size increases further. Investigating the interplay between reasoning diversity and model capacity at larger scales is an important avenue for future research.

\section*{Acknowledgments}
The work was partially supported by NSF awards \#2442477, \#2536297, \#2550203, \#2613637, \#2616632 and \#2623317. The views and conclusions in this paper should not be interpreted as representing any funding agencies.


\bibliography{acl_latex}

\appendix

\clearpage

\section{Theorem Prover: Prover9}
\label{app:prover9}

For \textsc{True} instances, the original conclusion $C$ serves as the conjecture, so a successful refutation establishes $\mathrm{Premises} \models C$. For \textsc{False} instances, we instead use $\neg C$ as the conjecture; a successful refutation then demonstrates $\mathrm{Premises} \models \neg C$. In both cases, the prover returns a refutation trace of the same structural form. For \textsc{Uncertain} instances, neither $C$ nor $\neg C$ is derivable from the premises. We verify this by employing a finite model builder, Mace4, to construct one model satisfying $\mathrm{Premises} \cup \{C\}$ and another satisfying $\mathrm{Premises} \cup \{\neg C\}$. The existence of both models confirms that the premises do not determine the truth value of $C$.

\section{Algorithm}
\label{sec:algorithm}
In Algorithm~\ref{alg:equivaug}, we show the pseudo code for \ours.

\begin{algorithm}[t]
\caption{\ours}
\label{alg:equivaug}
\begin{algorithmic}[1]
\Require Seed set $\mathcal{D}$ of (problem, solution) pairs; rule sets $\mathcal{R}_{\mathrm{math}}$, $\mathcal{R}_{\mathrm{logic}}$; domain $\delta \in \{\mathrm{math}, \mathrm{logic}\}$
\Ensure Augmented dataset $\mathcal{D}'$
\State $\mathcal{D}' \leftarrow \emptyset$
\For{each $(p, s) \in \mathcal{D}$}
  \State Parse $s$ into symbolic representation (SR) $\phi$
  \State $\mathcal{S} \leftarrow \emptyset$
  \Repeat
    \State Sample rule $r \in \mathcal{R}_\delta$ and apply $r$ to $\phi$ to get $\psi$
    \If{$\psi \not\equiv \phi \land \psi \notin \mathcal{S}$}
      \If{\textsc{Verify}$(\psi)$}
        \State $\mathcal{S} \leftarrow \mathcal{S} \cup \{\psi\}$
      \EndIf
    \EndIf
  \Until{iteration bound reached}
  \For{each $\psi \in \mathcal{S}$}
    \State Render $\psi$ to text
    \State $\mathcal{D}' \leftarrow \mathcal{D}' \cup \{(p, \mathrm{render}(\psi))\}$
  \EndFor
\EndFor
\State \Return $\mathcal{D}'$
\end{algorithmic}
\end{algorithm}

\section{Experiment Details}

\subsection{Configurations}
\label{app:config}

\paragraph{Supervised Fine-Tuning.}
All SFT experiments use the Hugging Face Transformers library. We apply full-parameter fine-tuning for Qwen2.5-1.5B and Qwen2.5-3B, and LoRA~\citep{hu2022lora} for Llama-3.1-8B. The two settings differ only in learning rate: full fine-tuning uses 2e-5, while LoRA uses 2e-4. All other hyperparameters are shared: AdamW optimizer \citep{loshchilov2017decoupled} with weight decay 0.01, cosine learning rate schedule with 100 warmup steps, per-device batch size of 16, and BF16 mixed precision. We train for 3 epochs on all datasets. For all methods except for the ground-truth, we train models on approximately 30,000 samples for GSM8K and 8,000 for FOLIO to ensure a fair comparison between \ours{} and other methods.

\paragraph{Reinforcement Learning (GRPO).}
We adopt GRPO~\citep{shao2024deepseekmath} implemented via verl 0.8.0dev0~\footnote{\url{https://github.com/volcengine/verl}} with vLLM \footnote{\url{https://github.com/vllm-project/vllm}} 0.11.2 as the inference backend. All RL experiments are conducted on 2 Nvidia A100-80G GPUs using SDPA attention. The reward signal is binary, exact-match correctness of the final answer. We use a learning rate of 1e-6, KL penalty coefficient $\beta = 0.0005$, sampling temperature of 0.7, 8 rollouts per prompt, and BF16 precision. We use 3 epochs for GSM8K and 10 epochs for Folio. For all methods, we use the original training set from both datasets as the training set for GRPO.

Adopted from metamath~\cite{yu2024metamath}, we use the following training prompt for both SFT and RL:

\begin{tcolorbox}[breakable,enhanced,
  colback=blue!3!white,
  colframe=blue!70!black,
  fonttitle=\bfseries,
  title={Training Template for SFT and RL}]

{\footnotesize\color{black}\begingroup
\setlength{\parindent}{0pt}
\obeylines

Below is an instruction that describes a task. Write a response that appropriately completes the request.

\#\#\# Instruction:
\{\{ message['content'] \}\}

\#\#\# Response:

\endgroup}

\end{tcolorbox}

\subsection{Metric: RPD}
\label{sec:app_rpd}

To quantify the step-level diversity between reasoning paths, we adopt the Reasoning Path Divergence (RPD) metric~\cite{ju2025reasoning}. Given two solutions $S_A$ and $S_B$, RPD first uses an LLM to decompose each into sequences of core logical step summaries: $S_A \rightarrow \{a_1, a_2, \ldots, a_m\}$ and $S_B \rightarrow \{b_1, b_2, \ldots, b_n\}$, where each $a_i$ or $b_j$ is a concise semantic description of a single reasoning step, abstracting away surface-level narrative and computational details. All step summaries are then encoded into vector representations $\mathbf{e}(\cdot)$ via an embedding model. Without loss of generality, let $m \leq n$. For each step $a_i$ in the shorter solution, RPD computes its minimum cosine distance to all steps in the longer solution, and the final score averages these distances:
{\small
\begin{equation}
    \mathrm{RPD}(S_A, S_B) = \frac{1}{m} \sum_{i=1}^{m} \min_{j=1,\ldots,n} \left(1 - \frac{\mathbf{e}(a_i) \cdot \mathbf{e}(b_j)}{\|\mathbf{e}(a_i)\| \, \|\mathbf{e}(b_j)\|}\right).
\end{equation}
}
This asymmetric design that always matches from the shorter to the longer solution ensures that a genuinely novel reasoning step is recognized as distant even when the remaining steps are subsumed by the longer solution. A higher RPD score indicates greater strategic divergence between two reasoning paths. In this paper, following the setting in the original RPD paper, we use Qwen-3-embedding-8B~\cite{yang2025qwen3} as the embedding model.

\subsection{Prompt Template}
\label{app:prompt}

\begin{tcolorbox}[breakable,enhanced,
  colback=blue!3!white,
  colframe=blue!70!black,
  fonttitle=\bfseries,
  title={Prompt Template for Translating GSM8K to Symbolic Presentation}]

{\footnotesize\color{black}\begingroup
\setlength{\parindent}{0pt}
\obeylines

\textbf{System Prompt:}

You convert a natural-language math solution into exactly ONE arithmetic expression.
Output rules:
Output ONLY a single expression. Do not output any text, explanation, equals sign, numeric result, or markdown/code fences.
Allowed characters: digits 0-9, decimal point '.', operators '+-*/', and parentheses '()' only. No variables, no units, no commas, no words.
If the answer contains annotated intermediate steps of the form <<expr=result>>result (or similar like expr=<<expr=result>>result), then reconstruct the final expression by replacing numeric intermediate results with their original expressions recursively. Wrap each replacement in parentheses when substituting to preserve order.
If the answer shows intermediate calculations without explicit <<...>> annotations (for example 3*2=6 then 6*2=12 then 12*52=624), infer the chain and expand likewise, replacing each derived number with its generating expression.
Do not evaluate or simplify numeric subexpressions. Keep the original operations. For fractions prefer the original fraction notation rather than its decimal form.
If an intermediate number has no known generating expression in the answer, leave that number as-is.
Keep parentheses to preserve the intended order of operations. Add parentheses around substituted subexpressions to avoid ambiguity.
If you cannot reconstruct an expanded expression, output a single expression built from the final arithmetic line as-is (still obey rules 1 and 2).
Use deterministic settings (temperature=0) and return a single-line expression.

\medskip

\textbf{User Prompt:}

For example, given the natural-language answer:
He writes each friend 3*2=<<3*2=6>>6 pages a week. So he writes 6*2=<<6*2=12>>12 pages every week. That means he writes 12*52=<<12*52=624>>624 pages a year.
The correct final arithmetic expression is:
((3*2)*2)*52

Natural-language answer:
\{answer\}

Return only the single final arithmetic expression that expands intermediate results into their original subexpressions (see system rules).

\endgroup}

\end{tcolorbox}

\begin{tcolorbox}[breakable,enhanced,
  colback=blue!3!white,
  colframe=blue!70!black,
  fonttitle=\bfseries,
  title={Prompt Template for Translating Back GSM8K to Natural Language}]

{\footnotesize\color{black}\begingroup
\setlength{\parindent}{0pt}
\obeylines

\textbf{User Prompt:}

Your task is to write a natural language solution for a math word problem based on a given algebraic expression.
The solution should:
1) Imitate the style, tone, and level of detail of the original natural language solution, but do not use any special formatting such as << >> or dollar signs.
2) Include the final line starting with '\#\#\#\#' and the numeric answer.
3) Do NOT simply copy the original solution. Adjust the order and grouping of computations so that they match the NEW expression.
4) Only output the new natural language solution, with no extra comments or explanations.

Here is the math problem, its ORIGINAL algebraic solution expression, and its original natural language solution:

[Problem]
\{question\}

[Original Expression]
\{expr\_original\}

[Original Natural Language Solution]
\{answer\_text\}

Here is the NEW algebraic solution expression that is mathematically equivalent and still solves the SAME word problem:

[New Expression]
\{expr\_variant\_clean\}

Write your new natural language solution below:

\endgroup}

\end{tcolorbox}

\begin{tcolorbox}[breakable,enhanced,
  colback=blue!3!white,
  colframe=blue!70!black,
  fonttitle=\bfseries,
  title={Prompt Template for Translating FOLIO Proofs Back to Natural Language}]

{\footnotesize\color{black}\begingroup
\setlength{\parindent}{0pt}
\obeylines

\textbf{System Prompt:}

You are an expert at explaining logical reasoning in clear, natural English. You translate formal logical proofs into plain-language explanations that read like a thoughtful person working through the problem. Vary your style — sometimes start with the conclusion and work backward, sometimes build up from the premises, sometimes highlight the key insight first. Do NOT use rigid formatting like `Step 1:' or `Step 2:'. Write flowing prose, as if explaining to a smart friend.

\medskip

\textbf{True User Template:}

Below is a logical reasoning problem. The premises logically entail the conclusion.
I will show you the premises (in both natural language and formal logic), the conclusion, and a formal proof from an automated theorem prover.

Your job: write a natural language explanation of WHY the conclusion follows from the premises. Trace the reasoning in your own words, referencing the key logical steps. End your explanation with a clear verdict line: `Therefore, the answer is: True'

[Premises - Natural Language]
\{record.get('premises\_nl', '(not available)')\}

[Premises - Formal Logic]
\{record.get('premises\_fol', '')\}

[Conclusion]
\{record.get('conclusion\_fol', '')\}

[Formal Proof (Prover9)]
\{record.get('solution', '')\}

Write your natural language explanation below. Be natural and varied in style — do not mechanically list steps. End with `The answer is: True'

\medskip

\textbf{False User Template:}

Below is a logical reasoning problem. The premises logically CONTRADICT the conclusion — the negation of the conclusion can be proved from the premises.
I will show you the premises, the conclusion, and a formal proof that the negation holds.

Your job: write a natural language explanation of WHY the conclusion is false given the premises. Show how the premises lead to the opposite of what the conclusion claims. End your explanation with: `Therefore, the answer is: False'

[Premises - Natural Language]
\{record.get('premises\_nl', '(not available)')\}

[Premises - Formal Logic]
\{record.get('premises\_fol', '')\}

[Conclusion]
\{record.get('conclusion\_fol', '')\}

[Formal Proof (negation proved)]
\{record.get('solution', '')\}

Write your natural language explanation below. Be natural — do not mechanically list steps. End with `The answer is: False'

\medskip

\textbf{Uncertain User Template:}

Below is a logical reasoning problem. The conclusion is INDEPENDENT of the premises — it can be neither proved nor disproved from what is given. There exist scenarios where the premises are true and the conclusion holds, and other scenarios where the premises are true but the conclusion fails.

Your job: write a natural language explanation of WHY we cannot determine the truth value. Point out what information is missing or why the premises don't constrain the conclusion. End your explanation with: `Therefore, the answer is: Uncertain'

[Premises - Natural Language]
\{record.get('premises\_nl', '(not available)')\}

[Premises - Formal Logic]
\{record.get('premises\_fol', '')\}

[Conclusion]
\{record.get('conclusion\_fol', '')\}

[Solver Output]
\{record.get('solution', '')\}

Write your natural language explanation below. Be natural — explain what's missing or ambiguous. End with `The answer is: Uncertain'

\endgroup}

\end{tcolorbox}

\section{More Experimental Results}
Here, we present more results on the accuracy trends and entropy.

\label{sec:more_experiment}
\subsection{Accuracy Trends}

\begin{figure}
    \centering
    \includegraphics[width=\linewidth]{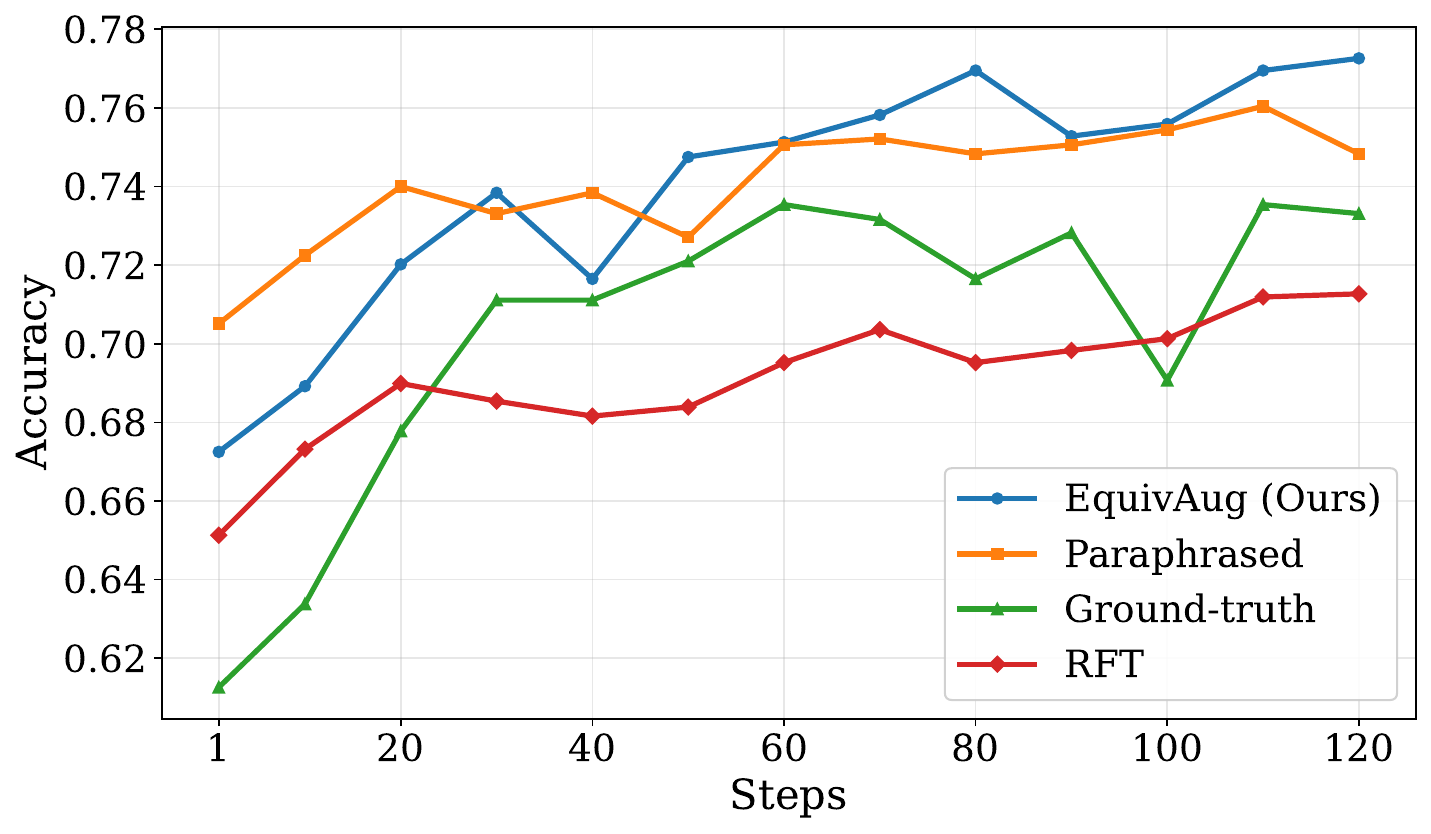}
    \caption{Test accuracy for different models during GRPO training for GSM8K on Llama3.1-8B.}
    \label{fig:accuracy_trend}
\end{figure}

In Figure~\ref{fig:accuracy_trend}, we show the evaluation accuracy trend during the GRPO training for GSM8K on Llama3.1-8B. The evaluation accuracy shows that even though \ours{} has lower initial accuracy, models trained on \ours{} can learn more effectively and outperform other methods soon.

\subsection{Entropy }

\begin{figure}
    \centering
    \includegraphics[width=\linewidth]{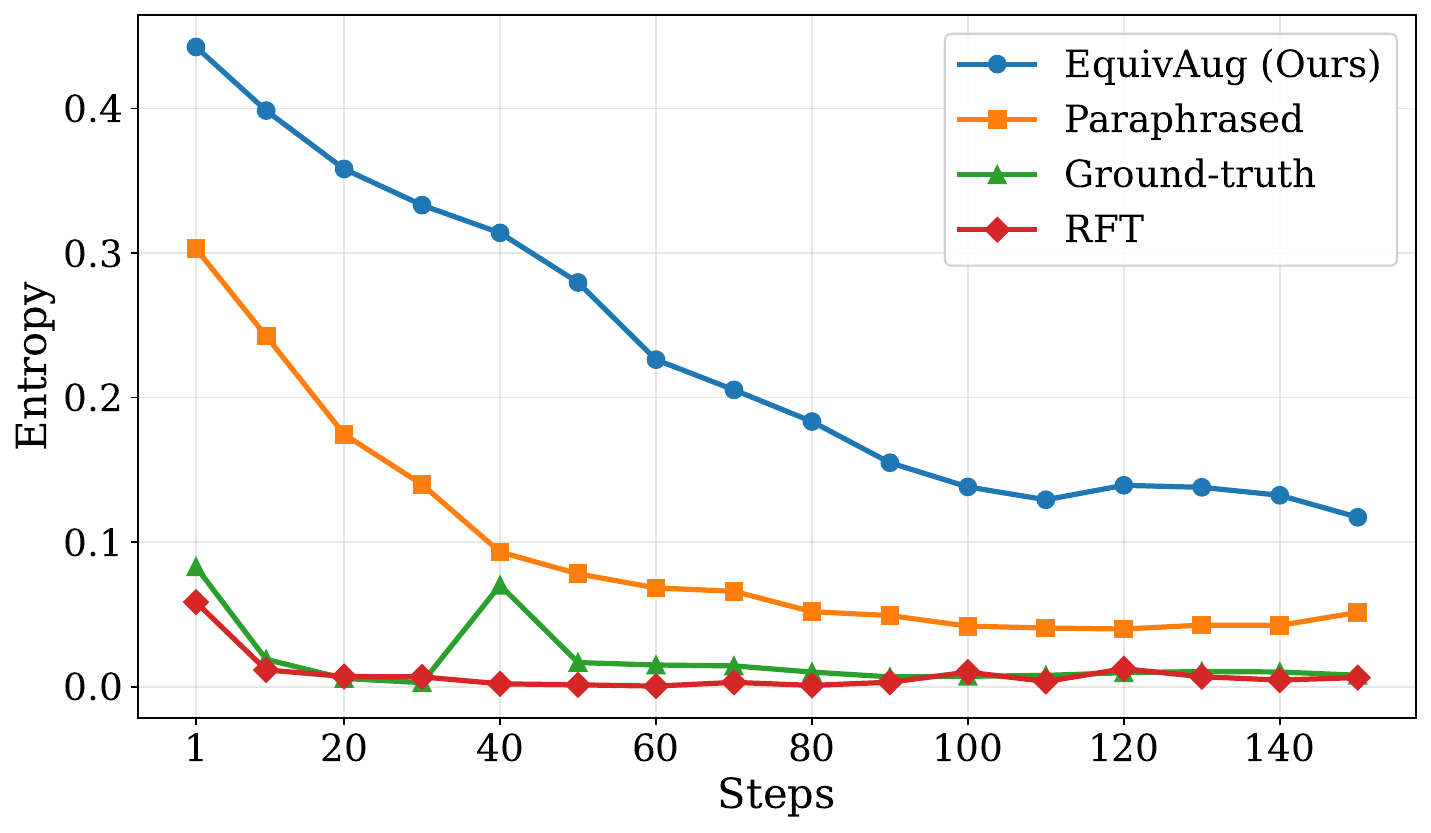}
    \caption{Entropy of different models during GRPO training for folio on Llama3.1-8B.}
    \label{fig:entropy}
\end{figure}

In Figure~\ref{fig:entropy}, we show how entropy changes for all methods on GSM8K and Llama3.1-8B.The results show that \ours{} has the highest entropy, further validating that \ours{} has a better exploration ability. Besides, the results show that with the training of GRPO, the entropy will decrease, which might be the reason that RFT has the lowest entropy since RFT can be seen as a weak version of RL without advantages and negative samples.

\label{sec:entropy}

\section{AI Assistant Usage}

We used GPT for grammar checking and Copilot as an assistive tool.
\label{sec:appendix}

\end{document}